# Mid-Infrared Optical Frequency Combs based on Difference Frequency Generation for Molecular Spectroscopy

Flavio C. Cruz[1,2,*], Daniel L. Maser[1,3], Todd Johnson[1,4], Gabriel Ycas[1], Andrew Klose[1], Fabrizio R. Giorgetta[1], Ian Coddington[1], and Scott A. Diddams[1,*]

[1] *National Institute of Standards and Technology, Boulder, Colorado 80305, USA*
[2] *Instituto de Fisica Gleb Wataghin, Universidade Estadual de Campinas, Campinas, SP, 13083-859, Brazil*
[3] *Department of Physics, University of Colorado Boulder, Boulder, Colorado 80309, USA*
[4] *Department of Physics, Saint John's University, Collegeville, Minnesota 56321, USA*
*\* [fcruz@nist.gov](fcruz@nist.gov); [sdiddams@nist.gov](sdiddams@nist.gov)*

**Abstract:** Mid-infrared femtosecond optical frequency combs were produced by difference frequency generation of the spectral components of a near-infrared comb in a 3-mm-long MgO:PPLN crystal. We observe strong pump depletion and 9.3 dB parametric gain in the 1.5 μm signal, which yields powers above 500 mW (3 μW/mode) in the idler with spectra covering 2.8 μm to 3.5 μm. Potential for broadband, high-resolution molecular spectroscopy is demonstrated by absorption spectra and interferograms obtained by heterodyning two combs.

## 1. Introduction

The development of femtosecond optical frequency combs (OFCs) in the mid-infrared (MIR) has been the subject of intense interest in recent years [1]. The principle motivation is the ability to access the strong vibrational fundamental bands of molecular gases in the so-called fingerprint region (2 μm - 20 μm). OFC spectroscopy has the potential to combine high-resolution and precision with broadband coverage, and may also allow real-time [2, 3, 4] and standoff detection [5]. These features are crucial for sensing and quantifying gases in a mixture, with obvious applications in environmental monitoring (greenhouse gases, pollutants) [6], security (hazardous gases) and defense (chemical weapons). Femtosecond OFCs in the MIR have been demonstrated using supercontinuum generation [7], optical parametric oscillators (OPOs) [8, 9, 10], difference frequency generation (DFG) [11, 12, 13], and fiber lasers [14]. There has also been some preliminary MIR comb work based on novel quantum cascade lasers (QCLs) [15, 16] and Kerr microcombs [17, 18].

Aside from traditional spectroscopic techniques [19], technology has been developed which exploits the advantages of OFCs, such as Vernier [20, 21], virtually-imaged phased array (VIPA) [22] and dual-comb spectroscopy (DCS) [23, 24, 25]. DCS brings significant advantages, including the use of a broadband coherent source, higher resolution and accuracy, and rapid scan rates without moving parts. Dual-comb, high-resolution spectroscopy has been demonstrated with DFG-based combs at 3 μm [26], but with low powers and narrow spectra. The technique has also been demonstrated recently with OPO-based combs [27, 8]. In Refs. [12, 13], MIR combs have been achieved via DFG between 1.0 μm and 1.5 μm pulses from Er fiber femtosecond combs, with MIR powers up to 120 mW and 150 mW, respectively. These sources were employed for absorption spectroscopy around 3 μm using grating-based or Fourier transform spectrometers.

Here we report MIR combs generated by DFG, which achieve powers in excess of 500 mW and whose spectra extend from 2.8 µm to 3.5 µm. This spectrum overlaps a relatively transparent window in the atmosphere and is well suited to observe fundamental bands of acetylene, methane, propane, ethane, and other hydrocarbons. The DFG has been characterized by strong pump depletion and significant parametric signal gain. Absorption spectra of acetylene and methane, recorded in grating spectrometers, along with interferograms obtained by heterodyning two combs, demonstrate good amplitude and phase noise properties of the combs, and therefore prospects for coherent, high-resolution dual-comb MIR spectroscopy at higher powers.

## 2. MIR DFG Optical Frequency Combs: Design and Characterization

The schematic diagram of one of the MIR DFG OFCs and setup for multiple heterodyne spectroscopy is presented in Figure 1, which is similar in design to those of Refs. [11, 12, 13, 28]. One advantage of DFG-generated OFCs is the cancellation of the carrier-to-envelope offset frequency ($f_{ceo}$) when the two wavelengths are coherently derived from the same femtosecond oscillator. The system starts with a home-built Er-doped fiber femtosecond oscillator based on nonlinear polarization rotation mode-locking, providing 20 mW of average output power at a repetition rate of 100 MHz. All subsequent components employ polarization-maintaining (PM) fibers and amplifiers, which were verified to provide better stability and robustness, as compared to our previous MIR comb generation based on non-PM fibers. The laser output is split into two branches with one being amplified in two Er-doped fiber amplifiers (EDFAs). The amplified signal beam extends from 1510 nm to 1625 nm with up to 140 mW of average power and pulses of 150 fs, measured using an autocorrelator and frequency-resolved optical gating (FROG) [29]. The other branch is spectrally broadened to produce a few milliWatts of light around 1050 nm (50 nm bandwidth) using 3 cm of PM highly nonlinear fiber (D = 5.6 ps/nm/km at 1550 nm) [30, 31]. These pulses are pre-amplified in a Yb-doped fiber amplifier (YDFA) and temporally stretched to near 50 ps using a few meters of fiber with negative third-order dispersion. The pulses are then sent to a high power YDFA, which uses a double-clad Yb PM fiber, pumped by two diode lasers (976 nm) with up to 8 W each, followed by a fiber-coupled optical isolator. A free-space pulse compressor with two transmission gratings produces a pump beam which extends from 1025 nm to 1070 nm, with 160 fs minimum duration pulses and average power up to 4 W. The signal (centered at 1567 nm) and pump (centered at 1048 nm) beams are spatially expanded and focused in a 3-mm-long MgO-doped periodically-poled lithium niobate (MgO:PPLN) crystal. A delay line stage is used to adjust the temporal overlap of the two pulses, but it has not been actively controlled.

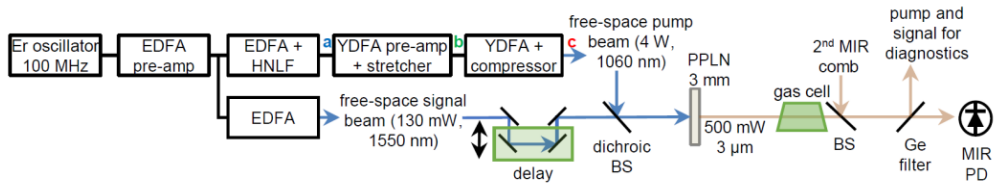

Fig. 1. Schematic diagram of the MIR DFG comb and heterodyne spectroscopy setup. Boxes are connected via fiber and the color lines represent free-space beams. Lenses (not shown) are used to expand the beams and focus them into the PPLN crystal. EDFA: erbium-doped fiber amplifier, YDFA: ytterbium-doped fiber amplifier, MIR PD: mid-infrared photodetector, Ge filter: germanium window, acting as a low pass optical filter, BS: beamsplitter, PPLN: periodically-poled lithium niobate crystal, HNLF: highly nonlinear fiber.

Our MgO:PPLN crystal is 3 mm long; AR-coated for the pump, signal and idler beams; and has five gratings with periods varying from 29.98 μm to 31.59 μm. Each one can generate DFG but with different central wavelength, spectrum and power. The PPLN crystal, placed in an oven, is operated near 150 °C. However, the MIR generation is not critically sensitive to temperature. Phase matching is achieved for e + e → e polarizations with effective nonlinear coefficient $d_{eff}$ = 14.9 pm/V [32]. Non phase-matched colors in the green (523 nm – second harmonic generation of the pump), red (628 nm – sum frequency generation: $f_{pump} + f_{signal}$), and UV (393 nm – $2f_{pump} + f_{signal}$; 349 nm – third harmonic generation of pump) are also generated, and the red beam is used as a guide for initial mode-matching and temporal overlap of the signal and pump pulses. This beam can also be useful for cross-correlation between the pump and signal pulses to estimate their duration. The pump and signal waist sizes in the crystal have been adjusted to near 50 μm and 70 μm, respectively, leading to confocal parameters larger than the crystal length, which is close to optimum. The beam waist sizes were chosen to keep the intensity below the manufacturer's stated damage threshold limit of about 4 GW/cm$^2$, and at the same time allow the use of most of the available pump power. A second MIR OFC was constructed with a similar design and is based on a commercial Er-doped fiber oscillator. The pump and signal beams are filtered out either after the PPLN or the gas cell using a germanium filter that transmits 90% of the MIR idler beam, which is then combined in a 50/50 CaF$_2$ beamsplitter with the beam from the other comb. The combined beams from one beamsplitter port go to a MIR Peltier-cooled pre-amplified HgCdTe photodetector (100 MHz bandwidth) and the other is used for diagnostics. Commercially available gas cells have been placed in the path of one of the MIR DFG combs for spectroscopy (Figure 1).

Figure 2 shows the MIR power of the first comb we have built as a function of pump power at 1050 nm for a signal input power of 130 mW. Over 500 mW, corresponding to roughly 3 μW/mode, is generated in the MIR, a power more than three times higher than what has been previously reported for DFG combs [12, 13]. Relatively good power stability has been verified for periods of a few hours, with a fractional standard deviation of 3.5% (Figure 2). The variation arises from temperature- and pressure-driven changes in the optical path lengths, neither of which are presently environmentally isolated. A similar plot for our previous comb based on non-PM fibers shows considerably higher power instability over a few minutes due to drift on the temporal overlap between the pump and signal pulses.

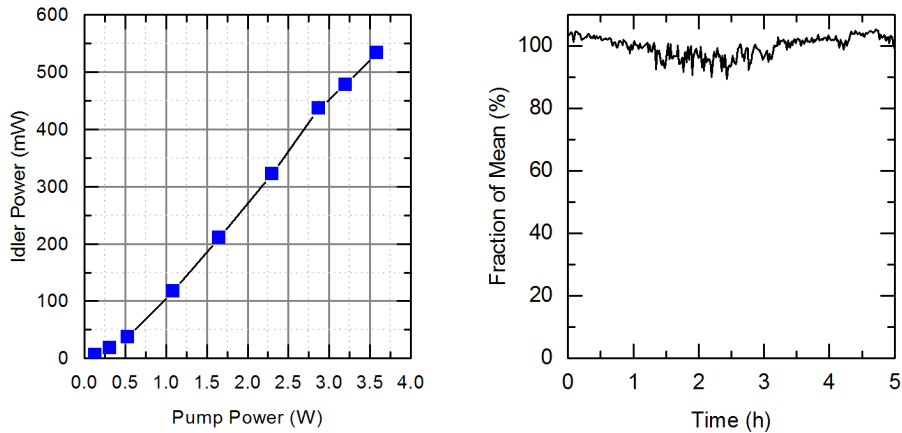

Fig 2. Left: MIR power (corrected for 90% transmission of the Ge filter) as function of pump power at 1050 nm, for a signal input power of 130 mW at 1567 nm. Right: MIR power stability plot, for PM-fiber-based MIR comb without environmental noise isolation.

Figure 3 shows the relative intensity noise (RIN) for the pump beam, measured along its fiber path in Figure 1. RIN from the oscillators, which can be increased by the several stages of amplification and via supercontinuum generation, is detrimental for dual-comb spectroscopy, degrading the SNR of the interferograms. The fiber lengths in the amplifiers have been optimized to minimize both the RIN and the pulse duration. The short pulse provides optimum broadening, maximizing the power at the relevant wavelengths. We have also verified that RIN for our previous comb based on non-PM fibers was higher.

An indirect measurement of the coherence of the combs has also been obtained by heterodyning both pump and signal pulses with narrow-linewidth ($\approx$ 1 kHz), single-frequency cw lasers at 1.0 µm and 1.5 µm. Beat notes with SNR near 40 dB and 10 kHz width have been obtained (recorded at RBW = 1 kHz and sweep time = 0.2 s), indicating good coherence for those beams after amplification.

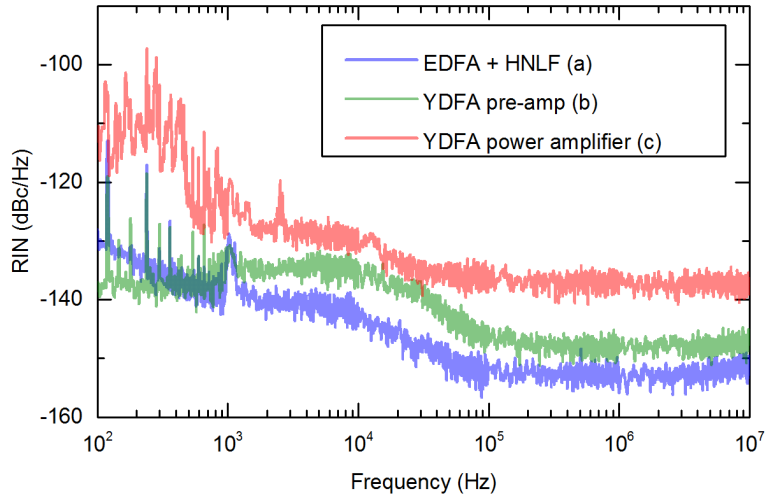

Fig 3. Relative intensity noise (RIN) for the pump beam, measured at positions along its fiber path in Figure 1. The letters (a), (b), and (c) denote the positions labeled in the schematic diagram in Figure 1.

Spectra obtained for one of the MIR combs are presented in Figure 4. Some absorption peaks seen in the spectrum on the left are due to atmospheric absorption in a 1.5 m path length. The spectra on the right show that the comb bandwidth is preserved as power is increased, therefore supporting the generation of ultrashort pulses. We changed the MIR idler power by changing the pump power with the current on the YDFA in Figure 1. The observed differences in the spectral envelopes on the right plot come from maximizing the power by readjusting the delay between signal and pump pulses, after changing the YDFA current.

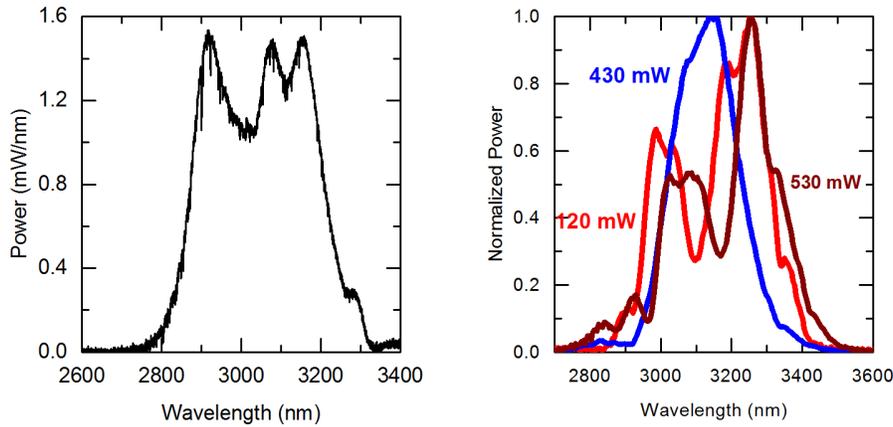

Fig 4. Left: MIR spectrum generated by DFG in a single PPLN grating, recorded in a grating-based optical spectrum analyzer (RBW: 0.2 nm). Absorption lines due to atmospheric propagation in a 1.5 m path length can be seen. Right: MIR comb spectra recorded in a lower-resolution FTIR spectrometer (RBW: 4 nm) for different powers, showing that the comb's bandwidth is preserved at high powers. Different spectra can be obtained as the temporal overlap between the pump and signal pulses is adjusted.

Figure 5 shows spectra for the pump and signal beams, measured after the PPLN both in presence and absence of nonlinear conversion, at the conditions of the maximum power curve in Figure 4. DFG is switched off either by blocking one of the beams or by adjusting the delay stage in order to avoid temporal overlap at the crystal. Given that the pump input spectrum extends approximately from 1025 nm to 1070 nm and the signal spectrum extends from 1510 nm to 1625 nm, the MIR spectral bandwidth should reflect their combined spectra, and thus would extend from 2770 nm to 3670 nm, which are the phase-matched wavelengths for one of our PPLN gratings (29.98 µm period). However, while the MIR spectra for all the PPLN gratings start at the expected lower limit (as seen in Figure 4), they extend only up to 3500 nm, indicating that DFG is less efficient on the red side.

The spectra in Figure 5 also show that DFG in the 3-mm-long PPLN crystal occurs with considerable pump depletion and high parametric gain for the signal beam. For example, a pump beam with 3.6 W has its power depleted to 2 W after the crystal (2.5 dB attenuation), while the signal beam is amplified from 130 mW to 1.12 W, corresponding to a power gain of 8.6 (9.4 dB or 7.2 cm$^{-1}$). These beams generate 530 mW of MIR power.

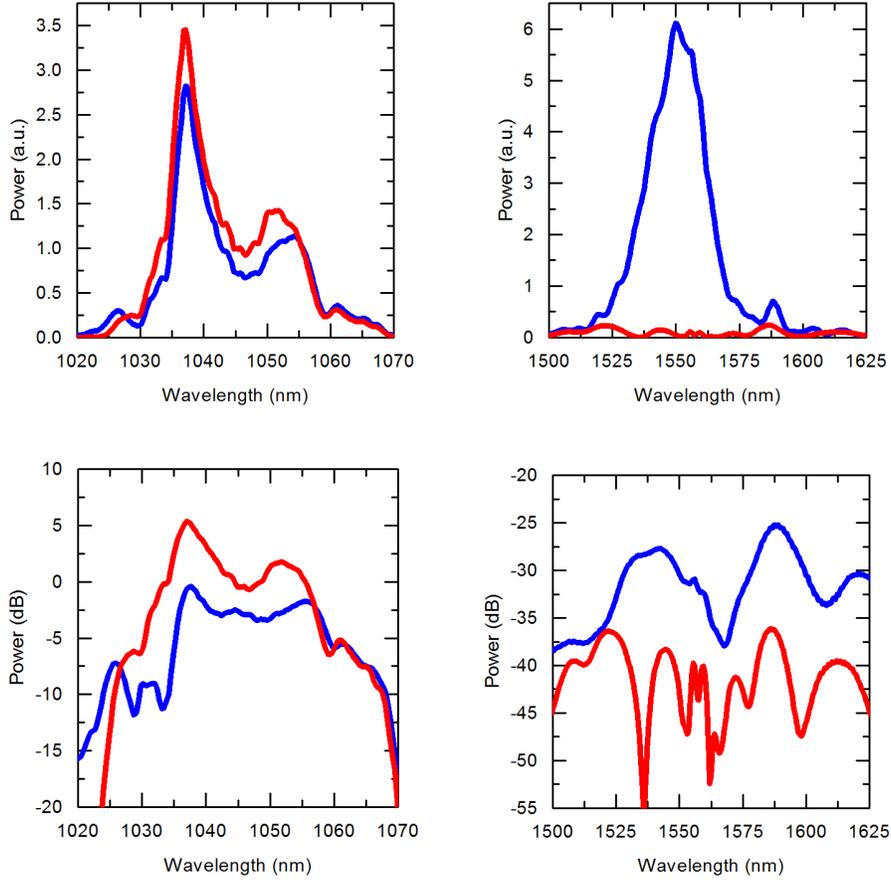

Fig 5. Pump and signal spectra measured after the PPLN crystal when both beams are present (DFG on, blue trace) and when one of them is blocked (DFG off, red trace). Four independent measurements are shown, in linear and log scale. The red side of the pump spectrum and the blue side of the signal spectrum undergo less depletion and amplification, respectively. The DFG is correspondingly less efficient on the red side of the idler spectrum.

A simplified analysis of the DFG, which does not initially take into account the broad spectral bandwidth of the pulses, was performed by solving the coupled equations for the pump, signal, and idler field amplitudes ($A_p$, $A_s$, $A_i$) [33]. The coupled equations, in which we consider the peak field amplitude of the pulses, are given as

$$\frac{dA_p}{dz} = \frac{8\pi\, i\, d_{\text{eff}}\, \omega_p^2}{k_p\, c^2} A_s A_i e^{-i\Delta k z},$$

$$\frac{dA_s}{dz} = \frac{8\pi\, i\, d_{\text{eff}}\, \omega_s^2}{k_s\, c^2} A_p A_i^* e^{i\Delta k z}, \text{ and}$$

$$\frac{dA_i}{dz} = \frac{8\pi\, i\, d_{\text{eff}}\, \omega_i^2}{k_i\, c^2} A_p A_s^* e^{i\Delta k z},$$

where $d_{\text{eff}}$ is the effective nonlinear crystal coefficient, $k_j = 2\pi/\lambda_j$, $\omega_j = k_j c/n_j$, and $\Delta k$ is the phase mismatch [33]. The calculated average powers, which can be compared to the measured ones, are related to the peak field amplitudes by $P_{avg} = \frac{n\,c}{2\pi}|A|^2(\pi\omega_0^2)\,t_{pulse}\,f_{rep}$, where $\omega_0$ is the laser waist size at the crystal, $t_{pulse}$ is the pulse duration, and $f_{rep}$ is the repetition rate. Figure 6 shows the calculated average powers for pump, signal, and idler, obtained by assuming the experimental parameters of the upper plots in Figure 5: pump and signal central wavelengths $\lambda_p$ = 1048 nm, $\lambda_s$ = 1568 nm; pump and signal average input powers $P_{avg}^{pump}$ = 3.6 W, $P_{avg}^{signal}$ = 130 mW; repetition rate $f_{rep}$ = 100 MHz; pulse durations $t_p = t_s$ = 160 fs; waist sizes $\omega_{0p} = \omega_{0s}$ = 70 μm; polarizations e + e → e; PPLN nonlinear coefficient $d_{eff}$ = 14.9 pm/V [32]; and perfect phase-matching ($\Delta k$ = 0). We obtain a pump depletion of 3.28 dB, signal amplification of 10.3 dB, and idler output power of 647 mW. These numbers are higher than what we have measured, and in fact very close agreement would be obtained by assuming a small amount of phase mismatch (namely, $\Delta k$ = 6 cm$^{-1}$). However, as discussed below, other factors not taken into account, such as the spectral structure of the pulses or group velocity mismatch (GVM) between them, can account for the difference.

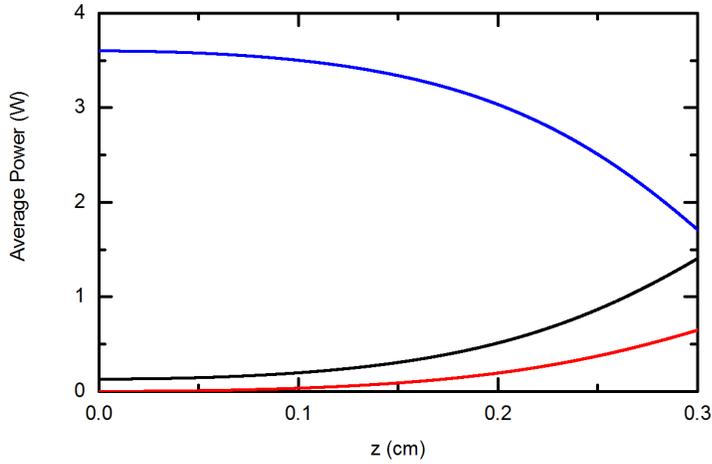

Fig 6. Simulated average powers for pump (blue), signal (black) and idler (red) beams as a function of propagating distance inside the 3-mm-long PPLN crystal.

From Figure 6, one can also ask if a longer crystal could be used to increase the idler power further. The DFG efficiency will be limited by GVM, which causes temporal walk-off between the pulses, and group velocity dispersion (GVD), which can broaden the pulses, reducing their intensities and the DFG efficiency. Both effects are characterized by effective lengths, which are useful to estimate the maximum crystal length over which GVM and GVD could be neglected. The GVM length is given by $L_{GVM} = t_p/GVM$, where $t_p$ is the pulse duration and GVM = $(1/v_{gp} - 1/v_{gs})$, where $v_{gp}$ and $v_{gs}$ are the group velocities for pump and signal pulses. Using $t_p$ = 160 fs, measured for the pump pulse, and the group velocities for the signal and pump beams for PPLN [32], we find $L_{GVM}$ = 1.47 mm. Therefore in our 3-mm crystal the pump-signal temporal walk-off is 2 × 160 fs, implying what is already the likely limit to more efficient MIR generation with the present crystal. The pump and idler pulses, which in turn amplify the signal, have more similar group velocities and could propagate together in the PPLN crystal over a GVM length of 3.8 mm, assuming 160 fs pulses. The dispersion length associated with GVD is given by $L_D = t_p^2/GVD$, where GVD = $d^2k/d\omega^2$ = 254 fs$^2$/mm for the pump pulses [32]. We then find $L_D$ = 10 cm at 1050 nm, indicating that a much longer crystal would still preserve the duration of the pump pulse, and therefore also its intensity and parametric gain. The dispersion lengths for the signal and idler pulses (also assumed to be 160 fs long) are 25 cm and 4.2 cm respectively, indicating that neither one is significantly broadened by propagation in the 3-mm crystal.

The poor conversion efficiency on the red side of the MIR spectrum that was discussed on Figures 4 and 5 may also be related to temporal walk-off due to GVM between the pulses. For example, if the pump and signal pulses have some amount of residual chirp and separate as they propagate into the crystal, a better temporal overlap could happen between higher frequencies of the pump pulse and the lower frequencies of the signal pulse, which generate higher frequencies in the idler pulse. On the other hand, lower frequencies of the pump pulse could quickly separate from the higher frequencies of the signal pulse, preventing the generation of lower frequencies in the idler spectrum. In such a case, spectral shaping of the input pump and signal pulses could be used to shape the MIR spectrum. However, this was not attempted and is beyond the scope of the present work.

### 3. Application to MIR Molecular Spectroscopy

The broad-bandwidth MIR spectra can be applied to the spectroscopy of molecular gases using a variety of detection techniques. As a first illustration, Figure 7 shows absorption spectra of the $\nu_3$ band of methane ($CH_4$) and acetylene ($^{12}C_2H_2$ and $^{13}C_2H_2$) in sealed gas cells, which were obtained directly from one of the MIR combs with a grating-based optical spectrum analyzer (OSA). These single-trace spectra, recorded with 0.2 nm resolution (0.2 cm$^{-1}$ or 6.0 GHz, comprising 60 comb modes), show good SNR and sensitivity, indicating a low comb intensity noise.

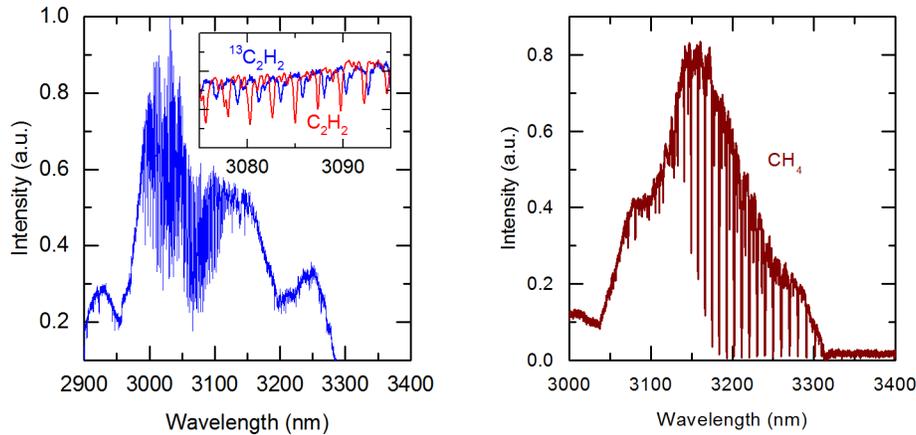

Fig 7. Left: Absorption spectra (without subtraction of the comb spectra) of two gas cells of $^{12}C_2H_2$ and $^{13}C_2H_2$ (75 mm long, 50 Torr). Inset: zoomed region shows isotope shifts of individual lines when each gas component is measured separately. Right: spectrum of a $CH_4$ cell (75 mm long, 200 Torr).

Ultimately, higher resolution spectra with the full amplitude and phase information of the absorbing media can be obtained via the multi-frequency heterodyne between two MIR combs. This requires high mutual coherence between the combs, which can be obtained by locking their repetition rate difference to a stable RF reference or by locking each comb to a common optical frequency reference. Even without fully implementing those techniques, we demonstrate good coherence between our MIR combs by recording the time-domain interferogram obtained by heterodying them. The left plot in Figure 8 shows an interferogram obtained with the setup of Figure 1 using a $C_2H_2$ cell (75 mm long, 50 Torr) in the path of one of the MIR combs. The interferogram is an average of five traces, and shows a central burst followed by the free-induction decay of the $C_2H_2$ molecules [34]. In this experiment, the repetition rate difference $\Delta f_{rep}$ between the combs was loosely stabilized to 96 Hz, using a low-bandwidth servo system to lock the repetition rate of one Er:fiber laser to a microwave signal synthesized from the second Er:fiber ("local oscillator"). For these data, the local oscillator is free-running. Its stabilization can be implemented in future experiments, but we note here that the offset frequency of the MIR comb is eliminated in the DFG process, such that absolute frequency information can be obtained with only $f_{rep}$ stabilized. The interferograms repeat at $(96\ Hz)^{-1} = 10.4$ ms, and a shorter 42 µs section is shown in Figure 8. The x-axis is the "laboratory time" as given by a fast acquisition oscilloscope, which records the output of the MIR photodetector, after passing through a 50 MHz low-pass filter. The optical power incident on the $C_2H_2$ cell was 50 mW, but an optical attenuator was placed before the MIR photodetector to avoid saturation by light from both combs.

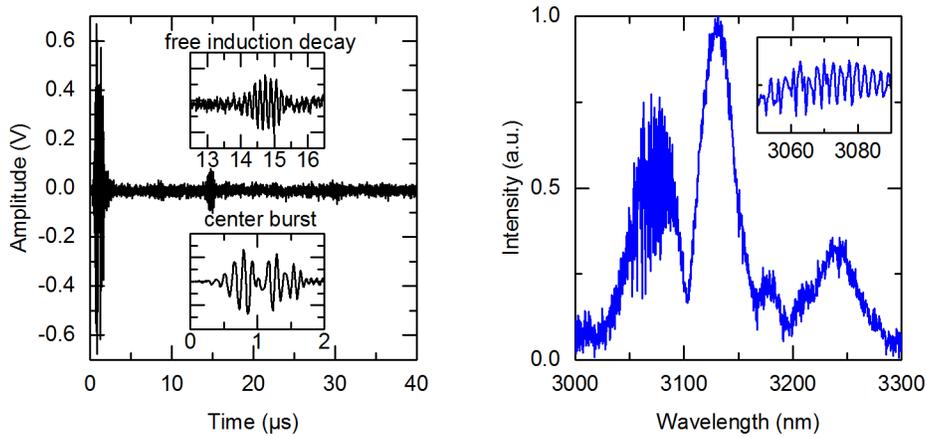

Fig 8. Left: Average of five interferograms from heterodyning two combs, with a $C_2H_2$ gas cell (75 mm long, 50 Torr) in the path of one of them. The central burst corresponds to the comb pulses crossing at the detector, and is followed by revivals due to molecular free-induction decay. The x-axis is the laboratory time. Right: Fast Fourier transform of the interferogram, revealing absorption lines of the $\nu_3$ band of $^{12}C_2H_2$.

The right plot in Figure 8 is an FFT of the interferogram, showing absorption lines of the $\nu_3$ band of $C_2H_2$. The spectral resolution is estimated to be 0.8 cm$^{-1}$ (25 GHz, or $\Delta\nu/\nu = 2.5\times10^{-4}$), limited by the 42 µs "lab time" duration of the interferogram, which corresponds to an "effective time" of 42 µs $\times$ $\Delta f_{rep}/f_{rep}$ = 40 ps. We are currently working to improve the SNR of the interferograms, characterize the phase noise of both combs, and improve the lock between them so that coherent averaging can be implemented for high-resolution spectroscopy [25].

## 4. Conclusion

In conclusion, 100 MHz MIR optical frequency combs based on PM fiber amplifiers and difference frequency generation (DFG) were developed. MIR spectra extended from 2.8 µm to 3.5 µm with average powers above 500 mW, corresponding to about 3 µW/mode. Use of PM fibers provided reduced amplitude noise compared to our previous comb based on non-PM fibers. MIR absorption spectra show good SNR and sensitivity. Simulations of the DFG power show good qualitative agreement with our experimental data. We demonstrate mutual coherence between two similar, weakly-locked combs with slightly different repetition rates, by examining the time-domain interferogram obtained from heterodyning them. An FFT of this interferogram reveals absorption lines of the $\nu_3$ band of acetylene. Future improvements in the lock of the MIR combs should allow longer averaging of the interferograms and dual-comb spectroscopy in the resolved comb mode regime. The high MIR average powers demonstrated here could also be useful for nonlinear optics (such as MIR supercontinuum generation), nonlinear molecular spectroscopy (such as two-photon), and engineering and coherent control of ro-vibrational states in molecules. Spectroscopy with simultaneous excitation at different spectral regions, naturally afforded by DFG combs, can be useful for molecular sensing in gas mixtures or in the presence of a complex matrix. Dual-band excitation (for example, at 3 μm and 1.5 μm) can possibly be explored also for quantum interference experiments involving fundamental and overtone bands, using 3-level schemes with rotovibrational transitions of molecular gases.

**Note:** After preparing this manuscript we noticed a recent publication [F. Zhu, A. Bicer, R. Askar, J. Bounds, A. A. Kolomenskii, V. Kelessides, M. Amani and H. A. Schuessler, Laser Phys. Lett. **12**, 095701 (2015)] in which the authors report MIR dual-comb spectroscopy of methane using similar DFG femtosecond fiber combs.

**Acknowledgments**
The authors thank William Loh and Richard Fox for comments on this manuscript, and also Nathan Newbury for discussions. We are grateful to M. Hirano of Sumitomo Electric Industries for providing optical fibers. We acknowledge the support of NIST (including the Greenhouse Gas and Climate Science Measurement Program) and the DARPA SCOUT program. FCC acknowledges Esther Baumann for discussions and loan of MIR optics, and also the support of CNPq, Fapesp, Fotonicom and UNICAMP. This paper is a contribution from the US government and is not subject to copyright in the US.